# Cybersecurity Career Requirements: A Literature Review

*Mike Nkongolo[1], Nita Mennega [1], Izaan van Zyl[1]
[1]University of Pretoria, Department of Informatics, Hatfield 0028, South-Africa

1**mike.wankongolo@up.ac.za
2nita.mennega@up.ac.za

* Mike Nkongolo

**Abstract.** This study employs a systematic literature review approach to identify the requirements of a career as a cybersecurity professional. It aims to raise public awareness regarding opportunities in the IS profession. A total of 1,520 articles were identified from four academic databases by searching using the terms "cybersecurity" and "skills". After rigorous screening according to various criteria, 31 papers remained. The findings of these studies were thematically analysed to describe the knowledge and skills an IS professional should possess. The research found that a considerable investment in time is necessary for cybersecurity professionals to reach the required technical proficiency. It also identified female gender barriers to cybersecurity careers due to the unique requirements of the field, and suggests that females may successfully enter at lower levels and progress up the tiers as circumstances dictate.

## 1 Introduction

A study conducted by the Cybersecurity Workforce revealed a three million professional shortage of cybersecurity professionals [1] due to the increase in cyber-threats [2–5]. In addition, hackers have more opportunities for security breaches due to the availability of the Internet connectivity as well as the Internet of Things (IoT). This has resulted in various issues such as data loss and financial impacts caused by ransomware [6]. A cyber-threat can also take the form of cyberbullying which harms an individual by psychologically embarrassing him. This may lead to the individual taking drastic steps. Cyber threats can also occur via home automation, where attackers use IoT to control home devices. One such example could take the form of a residence's gates being remotely controlled by a hacker with malicious intent [7]. Despite the shortage of cybersecurity professionals, females face various challenges in entering the cybersecurity space [8]. There are various risks involved in the job which causes the exclusion of women from the IS profession. These take the form of long working hours, where cybersecurity engineers are required to work on weekends and late at night for maintenance windows. Despite the risks and challenges that females face, there is still a demand for more women in the cybersecurity landscape.

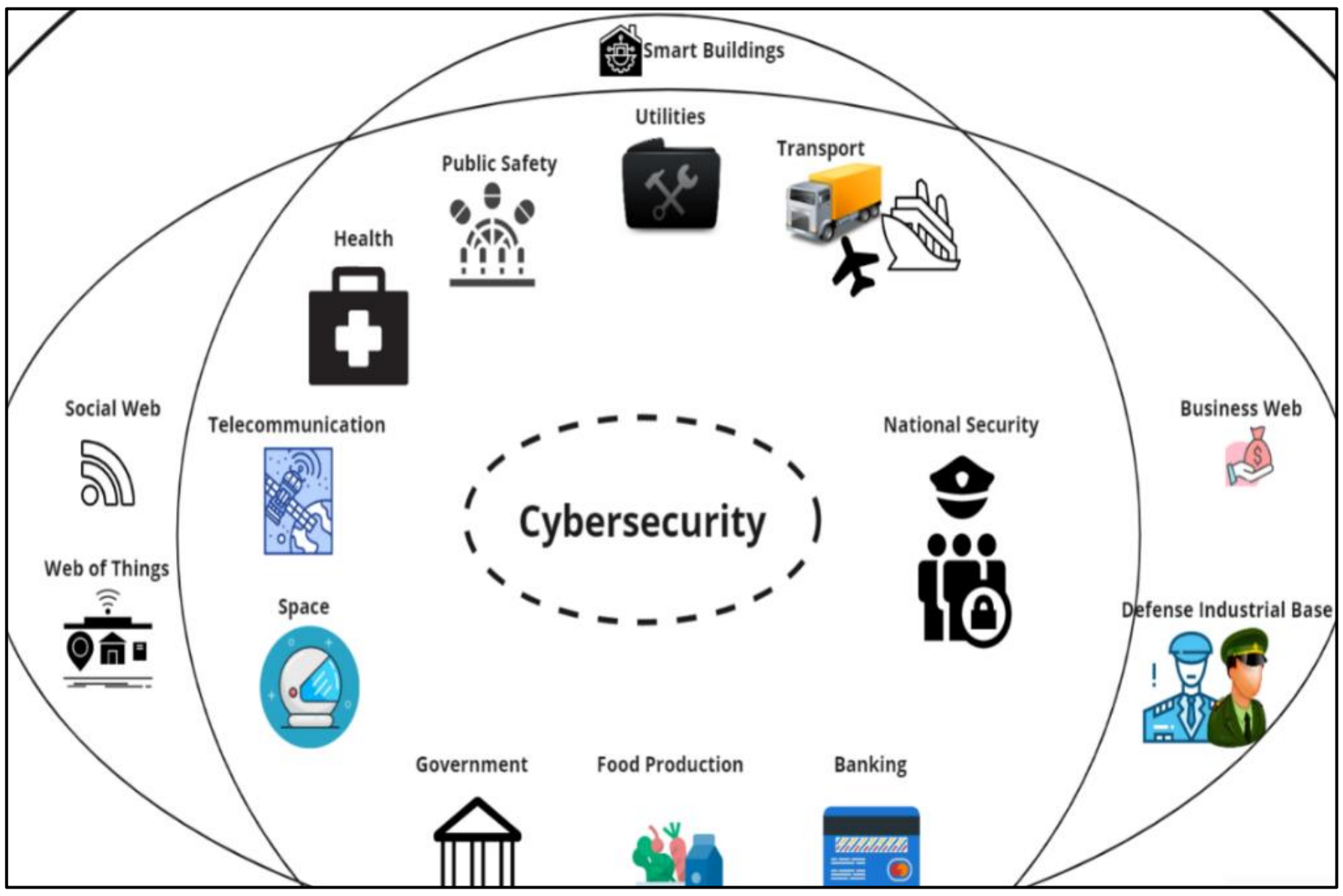

**Fig. 1.** The cybersecurity landscape (Source: Author)

Information Technology professionals should have an understanding of the hardware and software to implement and maintain cybersecurity systems (as illustrated in Figure 1). In essence, cybersecurity skills increase with experience and technological knowledge. Additionally, cybersecurity specialists need to possess specific personality traits such as attention to detail, scientific mindset, responsiveness, skepticism, and modesty [10]. The necessity of these traits stems from the job tasks which are usually performed under pressure. The systematic literature review aims to identify various demands placed on individuals pursuing a career in cybersecurity, as well as the barriers that may prevent them from pursuing a career in the IS field. However, the demands will vary depending on the job's specifications. The study discusses a cybersecurity awareness approach to encourage people to enter the IS field. The rest of this article is divided into the following sections: Section 2 provides a short background on the cybersecurity field; the research methodology is outlined in Section 3; and the results of the systematic literature review are introduced in Section 4. Section 5 concludes this study.

## 2 Cybersecurity Background

Many private and public enterprises are dealing with security issues related to the use of wireless communications [11]. These acts have serious consequences. They can disrupt government systems by installing a virus that locks confidential files and information, or malicious scripts may be sent to a country's power system instructing it to crash. These cyberattacks can disrupt the national security, banking, health, and space systems of any country, resulting in serious damage [11, 12]. Well-known techniques include vishing [13] which tricks people into disclosing private data such as social security numbers as well as bank account details. The top cybercrimes reported by the Federal Bureau of Investigation (FBI) since 2021 were phishing, scams (payment/non-delivery), investment fraud, and extortion [14]. Furthermore, the Internet Crime Complaint Center (IC3) received over 26,000 complaints concerning COVID-19 themes [15]. With problems such as these, IS skills are required in the field of cybersecurity to protect critical infrastructures from cyberattacks [16, 17].

### 2.1 Cybersecurity skills

Data privacy and security should be automated to protect critical infrastructures against an increasing number of threatening concerns. Network security schemes, backups recovery, policy enforcement, technical support, and systems monitoring are examples of cyber-security skills required to protect critical infrastructures [17, 18]. Other soft skills are crisis management, project management, and verbal or written communication. This study argues that only a few studies investigate cybersecurity careers in the IS landscape. For instance, [19] discovered the following IS skills: Problem-solving, the ability to develop positive customer relations, teamwork, leadership, stress management, negotiation, presentation, ethics, decision-making, and communication. In [20], employers were looking for the following technical skills: Experience related to cyber attack analysis, and skills in specific vendor products for network traffic management. Brooks and Greer [21] found that many IS-related jobs required work experience.

As a technical cybersecurity specialist, the duties might involve the following:

- Tier 1, Tier 2, and Tier 3 support for network Traffic Management Function (TMF) solutions.
- Monthly review of all client systems to identify issues that have to be addressed.
- Planning, coordinating, and implementing resolutions for identified issues on client systems.
- Performing monthly back-ups of each client's configuration [23, 24].
- Planning and initialising deployment and configuration of new trial master

file (TMF) solution elements.
- Performing client deployment, configuration, testing, and upgrades for in-house products.

In addition, the IS professional must be able to implement various graphical user interfaces for monitoring the network traffic to detect malicious intrusion. The interface should be intuitive and easy to use for novice operators, and powerful enough to enable more experienced users to see the level of detail necessary to effectively monitor and manage the traffic [23, 24]. The cybersecurity specialist must have a strong understanding of networking to configure different rules that control the network traffic, display statistics, monitor the detail of the traffic, manage system users, access system overview information, and view and manage system diagnostics to detect network attacks [24]. In some instances, the company will endeavor to provide the cybersecurity employee with the relevant training. The training might be provided in-house or when required, through different vendor partners' training programs. Unfortunately, there is often a substantial company capital outlay associated with these vendor training programs.

## 2.2 The work nature of a cybersecurity professional

A cybersecurity professional will focus on the Deep Packet Inspection (DPI) component of the critical infrastructure to detect network attacks. He/she will assist with the design, deployment, configuration, integration, and support of the network traffic management and related systems for a variety of clients [25]. The cybersecurity employee must apply data analysis and Machine Learning (ML) skills to create various forms of reports and dashboards that are used for operational purposes to detect malicious intrusion [26]. The cybersecurity professional must have troubleshooting skills to identify and resolve issues with the systems and the services provided over the network. Using insights combined with development and automation skills, he/she will assist in the continuous process of improving the team's productivity and the deployed system's reliability or performance. The cybersecurity engineer will endeavor to continuously improve the traffic management and related systems' efficiency, quality, and value to potentially detect network anomalies [27]. To be successful in this role, the candidate should possess a passion for exceptional quality with strong analytical thinking abilities and enjoy solving technical problems. The candidate must be innovative, and take the initiative to create solutions that improve intrusion detection.

## 2.3 Cybersecurity responsibilities

The responsibilities of a cybersecurity professional involve the following:

- Design, develop, and optimize traffic management systems in alignment with the client's requirements.
- Monitor the traffic management systems to proactively identify and resolve potential system issues.

- Troubleshoot technical issues.
- Ensure stability, performance, and end-to-end high availability/reliability of the deployed systems.
- Implement and test new and improved policies in the network traffic management system.

In addition, the cybersecurity technician must use knowledge and insights from the network traffic management and related systems to:

- Provide Tier 3 support, manage and resolve reported issues, and recommend and implement improvements to enhance the efficiency of network attack detection.
- Analyse statistics collected by the network traffic management system to build and analyze dashboards providing insights into the system performance, network utilization, subscriber behavior, and anomaly detection.
- Analyse data captured by the traffic management system / DPI and provide recommendations on improving the network services and quality of experience.
- Apply ML and predictive methodologies to enhance the accuracy of network attack detection.

To be a successful cybersecurity specialist, an individual should have good knowledge of networking protocols such as Internet Protocol (IP), Transmission Control Protocol (TCP), and the Open Systems Interconnection (OSI) model [28]. Strong knowledge of networked applications, such as SSL/TLS, HTTP, and Quic is important. The candidate should be comfortable working in a virtual environment having different Operating Systems such as Linux and Windows [29]. Working with time series data streams is advantageous as well as database administration and optimization with Structured Query Languages (SQL). Scripting and programming skills are important for data analysis, modeling, and interpretation of malicious intrusion. Dashboard creation and reporting will assist the cybersecurity professional in visualizing, categorizing, classifying, and predicting abnormal network concerns [2, 28, 29]. As such, a strong understanding of Artificial Intelligence (AI) and ML will be advantageous. Lastly, experience gained in the telecommunication sector is highly recommended for a successful cybersecurity career.

## 3 Research Methodology

A systematic literature review was utilized to locate relevant published research articles. Figure 2 illustrates the research questions.

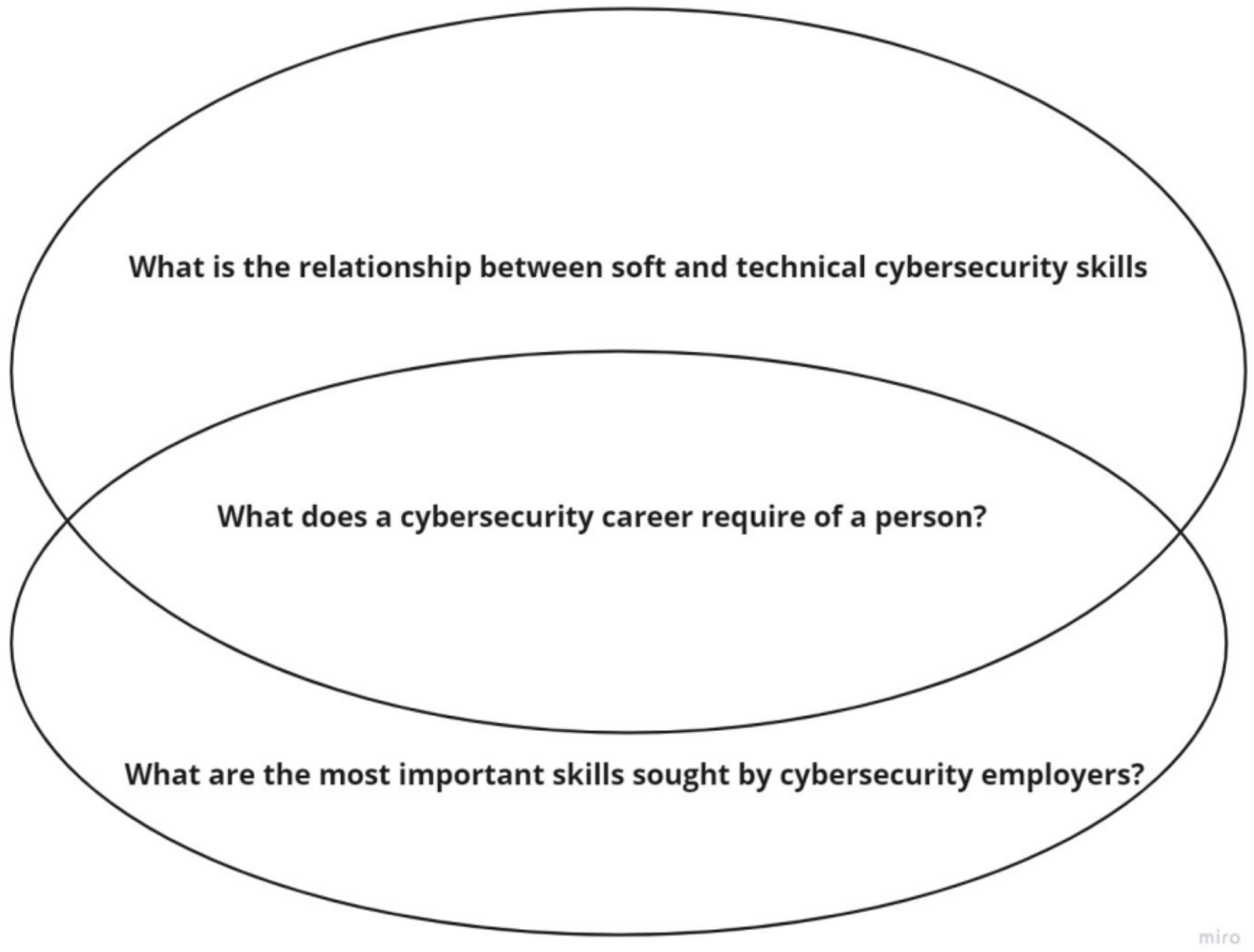

**Fig. 2.** The research questions

### 3.1 Keywords

The keywords shown in Figure 3 were utilized to search for relevant published articles.

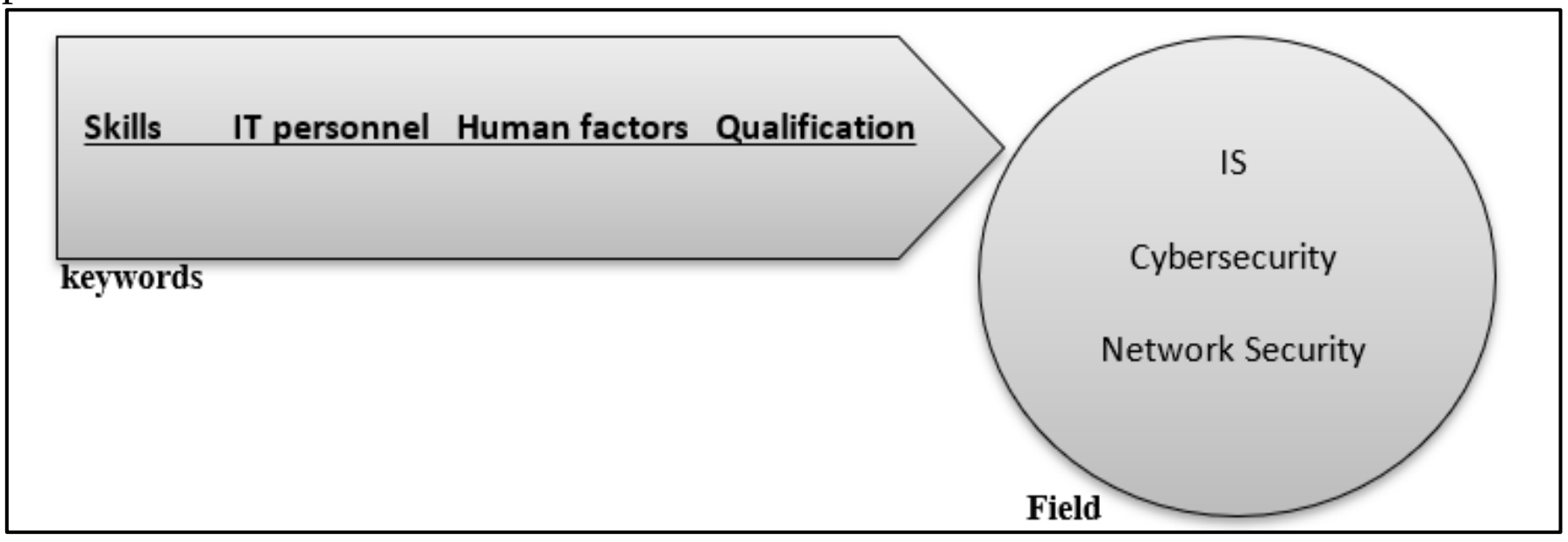

**Fig. 3.** The keywords used to identify relevant articles

### 3.2 Inclusion criteria

Academic articles and conference proceedings concerning cybersecurity skills, the cybersecurity work environment, and cybersecurity personality traits were included in this study.

### 3.1 Exclusion criteria

Books, theses, interviews, summaries, reviews, workshops, news articles, discussions, and preprints were all excluded from consideration here. Papers written in languages other than English were excluded from this study and so too papers that failed to describe the field of cybersecurity or the skills required therein.

### 3.1 Data collection

The following figure (Fig. 4) shows the databases used to collect data while Figure 5 illustrates the process of screening articles.

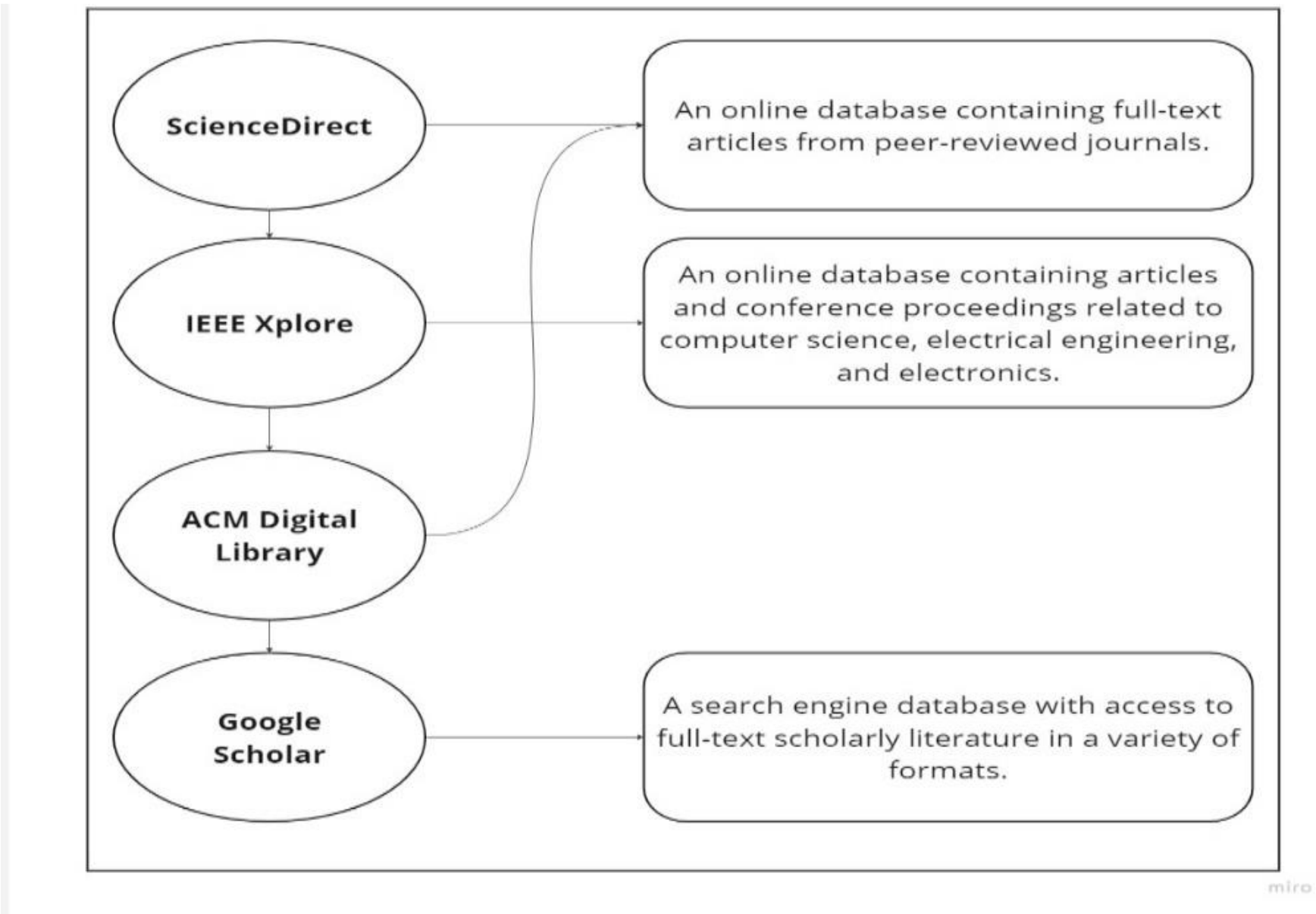

**Fig. 4.** Academic databases consulted

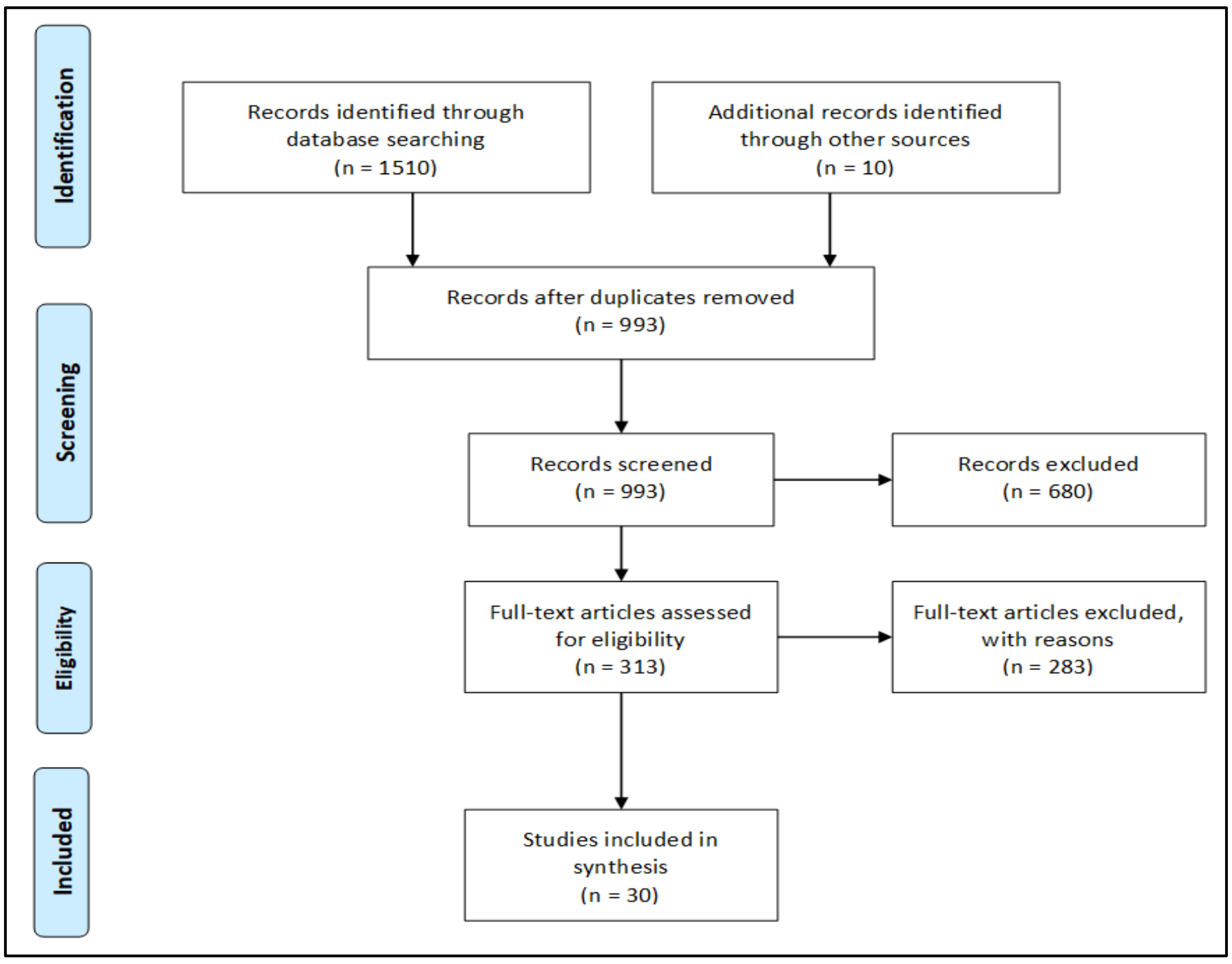

**Fig. 5.** The PRISMA flow diagram illustrating the screening process

Figure 5 contains the PRISMA diagram for this study. PRISMA is an acronym for "Preferred Reporting Items for Systematic Reviews and Meta-Analyses" which is a globally recognized standard for reporting evidence in systematic reviews as well as meta-analyses. These standards originated and are endorsed by journals and organizations in the health sciences. It allows authors to demonstrate the quality of their review, allow readers to assess its strengths and weaknesses, and permits replication of the review. After searching for literature using the abovementioned keywords (Figure 3), and downloading a total of 1,520 citations, the authors first removed any duplicates. They then screened the literature by perusing the title and abstract of the articles and considered if the contents would assist in answering the research questions. The same was subsequently done using the full text of each remaining article. The systematic search concluded with 31 relevant papers. The final papers are listed in Table 1. Figure 6 is a Wordcloud of the most important keywords retrieved from the cybersecurity literature.

**Table 1.** Final papers (Key: Aus - Australia, SA - South Africa)

| Nr | Citation | Country | Focus of paper |
|---|---|---|---|
| 1 | (Yair Levy, 2013) | USA | Skills needed for a career in cybersecurity. |
| 2 | (Von Solms & Van Niekerk, 2013) | SA | Threats and risks involved in the cyber environment and how it impacts individuals. |
| 3 | (Reeves et al., 2021) | Aus | Training is needed to acquire the appropriate skills for cybersecurity. |
| 4 | (Li et al., 2019) | USA | Training is needed to acquire the appropriate skills for cybersecurity. |
| 5 | (Hoffman et al., 2012) | USA | Training is needed to acquire the appropriate skills for cybersecurity. |
| 6 | (Hadlington, 2017) | UK | The focus is on the skills needed for a career in cybersecurity. |
| 7 | (Gratian et al., 2018) | USA | Intentions of an individual when working in a cyber domain. |
| 8 | (Furnell et al., 2017) | UK | Skills needed for a career in cybersecurity. |
| 9 | (Furnell, 2021) | UK | Skills needed for a career in cybersecurity. |
| 10 | (Dawson & Thomson, 2018) | USA | Workforce and the work environment of a cybersecurity specialist. |
| 11 | (Catota et al., 2019) | USA | Training needed to acquire the appropriate skills for cybersecurity. |
| 12 | (Caldwell, 2013) | USA | Skills needed for a career in cybersecurity. |
| 13 | (Burley et al., 2014) | USA | Workforce and the work environment of a cybersecurity specialist. |
| 14 | (Shahriar et al., 2016) | USA | Skills that need to be improved for a cybersecurity career. |
| 15 | (Jeong et al., 2019) | Aus | Personality of an individual working in a cybersecurity career |
| 16 | (Chowdhury et al., 2018) | Aus | Time pressure, how it impacts individuals with tasks to be completed in a certain timeframe. |
| 17 | (Paulsen et al., 2012) | USA | Workforce and the work environment of a cybersecurity specialist |
| 18 | (Bagchi-Sen et al., 2010) | USA | Skills needed for a career in cybersecurity |
| 19 | (Liu & Murphy, 2016) | USA | Education needed to acquire the appropriate skills for cybersecurity and which gender is mostly in the cybersecurity career. |
| 20 | (Javidi & Sheybani, 2018) | USA | Training needed to acquire the appropriate skills for cybersecurity. |
| 21 | (Sharevski et al., 2018) | USA | Training skills for cybersecurity. |
| 22 | (Besnard & Arief, 2004) | UK | Computer security and factors that influence cybersecurity. |
| 23 | (Martin & Rice, 2011) | Aus | Cybercrime and the impact it has on individuals and communities. |
| 24 | (Bauer et al., 2017) | USA | Making people and communities aware of the impact of cybercrime so that they are more careful with their information. |
| 25 | (Abawajy, 2014) | Aus | Make people aware of information security in different ways. |
| 26 | (Albrechtsen & Hovden, 2009) | Norway | Emphasize the different viewpoints between specialists and users of information security. |
| 27 | (Christopher et al., 2017) | USA | Identify the cybercrime trends and educate specialists. |
| 28 | (Baskerville et al., 2014) | USA, Italy | Identify and explain the threat paradigm versus the response paradigm of a cyber-attack. |
| 29 | (Rajan et al., 2021) | India, UK | Management of cybersecurity in an organisation. |
| 30 | (Hong & Furnell, 2021) | China, UK, SA | Behaviour of individuals exposed to cybercrime and how they respond to it. |
| 31 | (Kam et al., 2020) | USA | Learning required to improve skills of the cybersecurity workforce. |

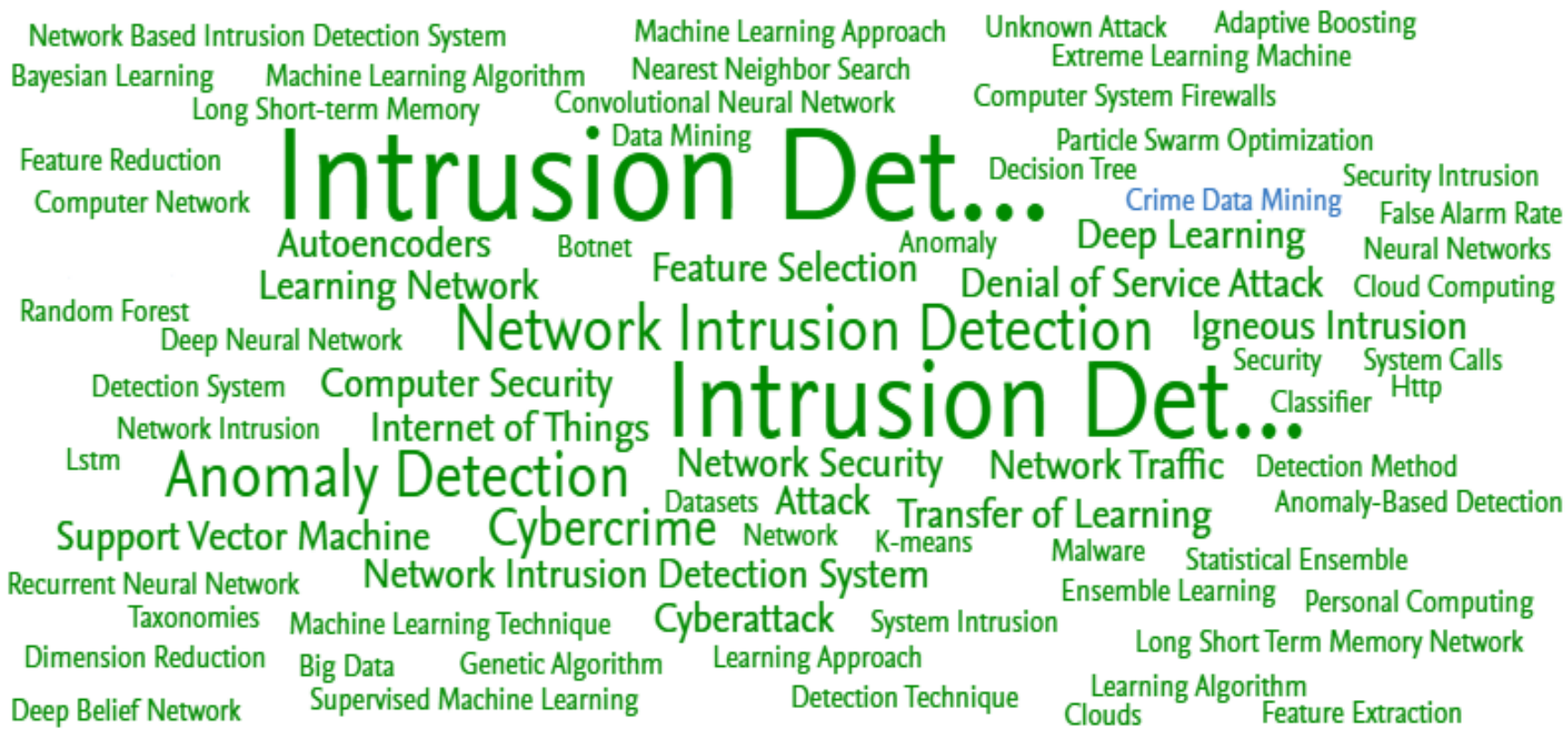

**Fig. 6.** The keywords used in the cybersecurity landscape

The Wordcloud shows that there is a growing application of machine learning (ML) in the IS industry. ML can be viewed as a new set of skills that cybersecurity professionals should possess. ML is a subset of artificial intelligence (AI) that refers to the process of teaching algorithms to learn patterns from existing data [30]. Although the terms AI and ML are frequently used interchangeably, there are significant differences between the two concepts. AI is the technology that trains machines to simulate human intelligence processes, whereas ML is the resulting computer systems (models) that learn from data to make predictions and classification [30,35]. In the cybersecurity space, ML has a diverse and ever-expanding set of applications. Those use cases can be classified into two categories: threat detection and threat response automation. Malware classification is one of the most common applications of ML in cybersecurity [2, 3, 30].

Generally, malware classifiers produce a scored prediction on whether a given network traffic sample is malicious or not. Here, the score refers to the level of confidence associated with the resulting classification or predictive process. The cybersecurity professional must have ML skills to evaluate these models' performance by plotting predictions along two axes: accuracy (whether an outcome was correctly classified; true or false) and output (the class a model assigns to a sample; positive or negative). The ultimate goal of developing high-performance ML models for cybersecurity is to improve the automation of network attack detection.

### 3.5 Quality assessment

The final 31 research articles were assessed with four quality assurance (QA) questions to judge the quality of each article. Each article was given a subjective score representing its relevance. The four QA questions are depicted in Figure 7. Each question is abbreviated and used as column headings in Table 2, which lists the QA ratings awarded to each paper.

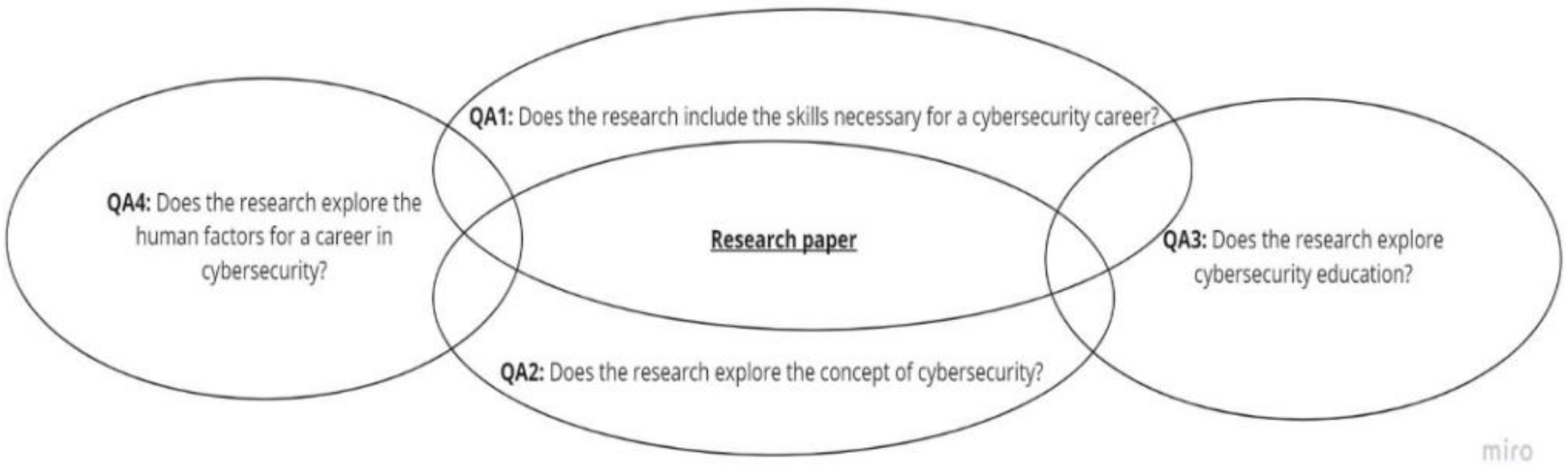

**Fig. 7.** The quality assurance (QA) questions

**Table 2.** The quality assurance (QA) ratings applied to the final articles

| Citation | QA1 | QA2 | QA3 | QA4 | Score |
|---|---|---|---|---|---|
| (Yair Levi, 2013) | Partial | Yes | No | Partial | 2.0 |
| (Von Solms & Van Niekerk, 2013) | No | Yes | No | Partial | 1.5 |
| (Reeves et al., 2021) | No | Yes | Yes | Partial | 2.5 |
| (Li et al., 2019) | No | Yes | No | Yes | 2.0 |
| (Hoffman et al., 2012) | Partial | Yes | Yes | No | 2.0 |
| (Hadlington, 2017) | No | Yes | No | Yes | 2.0 |
| (Gratian et al., 2018) | No | Yes | Partial | Yes | 2.0 |
| (Furnell et al., 2017) | No | No | Partial | No | 0.5 |
| (Furnell, 2021) | Yes | Yes | Partial | Partial | 3.0 |
| (Dawson & Thomson, 2018) | Yes | Yes | Partial | No | 2.5 |
| (Catota et al., 2019) | Yes | Yes | Yes | No | 3.0 |
| (Caldwell, 2013) | Partial | Yes | Partial | No | 2.0 |
| (Burley et al., 2014) | No | Yes | Partial | No | 1.5 |
| (Shahriar et al.) | No | Partial | No | No | 0.5 |
| (Jeong et al.) | No | Yes | Partial | Yes | 2.5 |
| (Chowdhury et al.) | No | Yes | Partial | Yes | 2.5 |
| (Paulsen et al., 2012) | Partial | Yes | Yes | No | 2.5 |
| (Bagchi-Sen et al., 2010) | Yes | Yes | Yes | Partial | 3.5 |
| (Liu & Murphy) | Yes | Yes | Partial | Partial | 3.0 |
| (Javidi & Sheybani) | Yes | Yes | Yes | Partial | 3.5 |
| (Sharevski et al.) | Partial | Yes | Yes | No | 2.5 |
| (Besnard & Arief, 2004) | No | Yes | Partial | No | 1.5 |
| (Martin & Rice, 2011) | No | Yes | No | No | 1.0 |
| (Bauer et al., 2017) | No | Yes | Partial | No | 1.5 |
| (Abawajy, 2014) | No | Yes | Partial | No | 1.5 |
| (Albrechtsen & Hovden, 2009) | Partial | Yes | No | No | 1.5 |
| (Christopher et al., 2017) | No | Yes | Partial | Partial | 2.0 |
| (Baskerville et al., 2014) | No | Yes | No | No | 1.0 |
| (Rajan et al., 2021) | No | Yes | Partial | No | 1.5 |
| (Hong & Furnell, 2021) | No | Yes | Partial | Yes | 2.5 |
| (Kam et al., 2020) | Partial | Yes | Yes | No | 2.5 |

Table 2 shows that four papers exhibit a QA rating score of 3 or more out of 4, and only three papers received a score of 1 or less. This indicates that most of the final 31 papers were highly relevant and potentially allow the authors to collate informative answers to the research questions. Thematic analysis was utilized in this respect. The next section provides some demographic information underlying the data and describes the results of the analysis.

## 4 Results

Figure 8 illustrates the number of cybersecurity research articles published since 2004. The chart reflects a steadily increasing interest in cybersecurity, including an unexpected dip during the year 2020 (the year where the world's working population had to revert to working online from home), but followed by a sharply renewed interest in the subject during the year 2021.

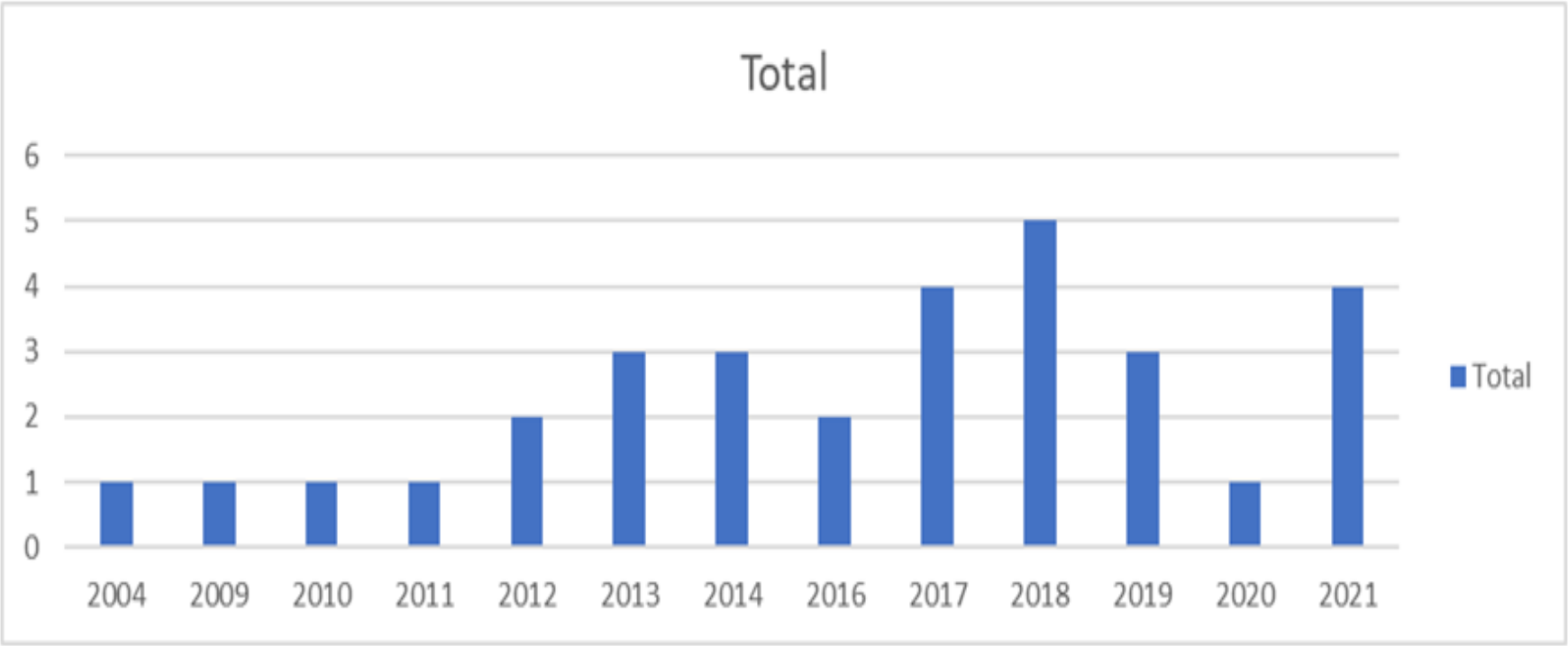

**Fig. 8.** Yearly count of cybersecurity articles

Figure 9 illustrates the number of articles that are published according to different cybersecurity concepts. As the column chart clearly illustrates, the most popular concept identified, was the skills required for a career in cybersecurity. The next two most popular topics pertained to cybersecurity education and training, and the cybersecurity workforce.

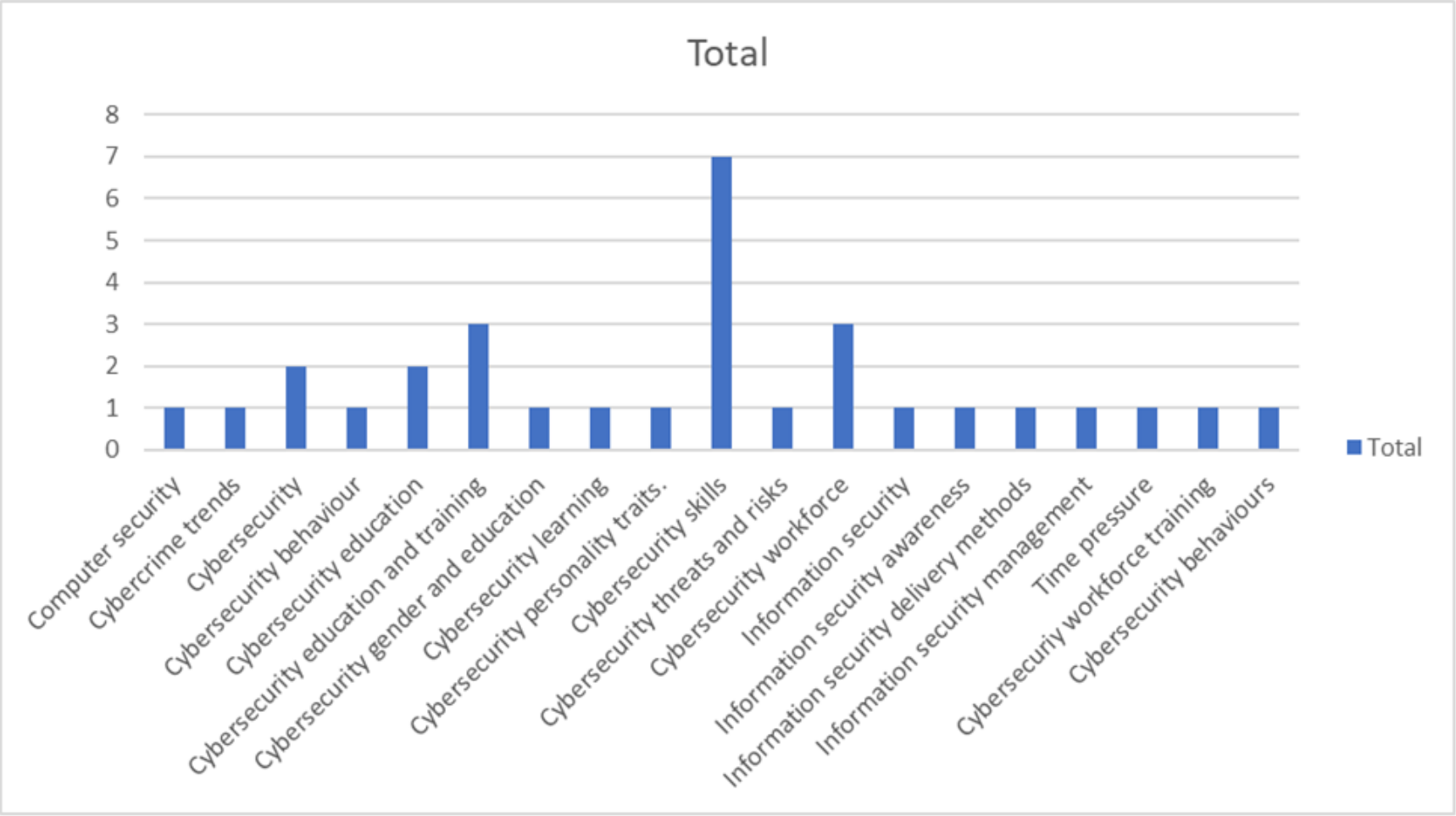

**Fig. 9.** Main concepts distilled from reviewed papers

This study discovered that Australia and the United States have the highest number of published articles about cybersecurity. Cybersecurity ensures an organization's data protection and privacy [24]. The main problem in the field of cybersecurity is cyber-attacks that compromise the security of critical infrastructures. In this light, cybersecurity professionals require specific skills to be successful [18]. Organizations can decrease the skills gap by creating cybersecurity awareness programs and training. Cybersecurity awareness programs can be delivered in various forms including paper, game, simulation, and video-based training programs. Organizations can make use of these different methodologies to create awareness about cybersecurity together with the requirement of a career in cybersecurity.

In addition to defining what a cybersecurity career requires of a person, this study also attempts to answer the following questions:

- What are the most important IS skills that employers look for? And, in terms of corporate expectations,
- How relevant are technical skills compared to soft skills?

The study found that more technical skills were required for a career in cybersecurity than soft skills. In [22], the ability to work independently was one of the most important soft skills recommended. Other important soft skills include knowledge management, project management, teamwork, and collaboration. In terms of technical skills, the knowledge of the vendor product, troubleshooting skills, and the ability to learn new technologies were identified. With regards to working hours, IS jobs in cybersecurity are full-time employments that require an approximation of 40 hours per week with some days requiring after-hour support to clients.

## 5 Conclusion

In this study, 31 articles were selected for a systematic literature review that provides evidence of what a career in cybersecurity requires of an individual, and what skills or knowledge are needed. The data collection was performed according to the systematic literature review methodology. The data was collected from bibliographic databases using specific keywords to answer the research questions. The results were thematically analysed to answer the research questions. According to the findings, the majority of articles investigate cybersecurity skills as well as the education, training, and awareness that must be created to alert the public about the impact of cybersecurity. This study also identified a demand for females in the cybersecurity field. The study found that different policies can be implemented for data security and privacy, but this depends on the company's maturity. In addition, game theory can also be used to develop strategic games that make cybersecurity professionals aware of the data protection policy. In the future, the authors intend to create a cybersecurity awareness game for data protection and privacy. Lastly, this study looked into the high demand for females in the cybersecurity field, who may be encouraged to enter the field at entry-level/Tier 1 support and then gradually gain experience while developing their industry qualifications.